\documentclass[11pt]{article}
\usepackage[pdftex]{graphicx}
\usepackage{wrapfig}
\usepackage{latexsym}
\usepackage[pdftex]{color}
\usepackage{pifont}
\usepackage{epstopdf}
\usepackage{tikz}
\usetikzlibrary{decorations.pathmorphing,patterns}

\newcommand{\bmp}[2][t]{\begin{minipage}[#1]{#2}}
\newcommand{\emp}{\end{minipage}}

\usepackage{amsmath}
\usepackage{amssymb}
\usepackage{textcomp}
\usepackage[square,numbers,sort&compress]{natbib}
\begin{document}
\title{Negative membrane capacitance of outer hair cells:\\
electromechanical coupling near resonance}
\author{Kuni H. Iwasa\\
35A Convent Dr., Rm 1F242A\\
NIDCD, National Institutes of Health\\
Bethesda, Maryland 20892
\footnote{email: kuni.iwasa@gmail.com}}
%\date{\small{Version 3: \today}}
\date{}

\maketitle

\subsection*{Abstract}
The ability of the mammalian ear in processing high frequency sounds, up to $\sim$100 kHz, is based on the capability of outer hair cells (OHCs) responding to stimulations at high frequencies. These cells show a unique motility in their cell body coupled with charge movement. With this motile element, voltage changes generated by stimuli at their hair bundles drives the cell body and that, in turn, amplifies the stimuli. \emph{In vitro} experiments show that the movement of these charges significantly increases the membrane capacitance, limiting the motile activity by additionally attenuating voltage changes. It was found, however, that such an effect is due to the absence of mechanical load. In the presence of mechanical resonance, such as \emph{in vivo} conditions, the movement of motile charges is expected to create negative capacitance near the resonance frequency. Therefore this motile mechanism is effective at high frequencies.

\textit{key words: mammalian ear, cochlear amplifier, piezoelectricity}

\subsection*{Introduction}
The remarkable sensitivity and frequency bandwidth as high as 100 kHz, depending on animal species  \cite{long1994}, of the mammalian hearing is based on the ability of its ear to function a frequency analyzer \cite{bekesy1960}. The frequency components are then transferred to the brain in parallel by a bundle of neurons. Thus a key question is how a cell-based biological system can be sensitive and capable of operating at high frequencies.

It has been found that an amplifier that counteracts viscous drag is essential for a sensitive mechanoeletrical analyzer such as the mammalian ear \cite{g1948,z1991} and that outer hair cells (OHCs) play such a key role \cite{ld1984,Dallos2008}. These cells have a motile mechanism in their cell body based on piezoelectricity, called ``somatic motility'' or ``electromotility,'' utilizing electrical energy \cite{a1987,sd1988,i1993,ga1994,doi2002}. The electric potential that is used by the motile mechanism is generated by mechanotransducer current of the sensory hair bundle of these cells, in response to mechanical stimuli. This process is assisted by the endocochlear potential, the unusual positive potential in the K$^+$-rich endolymphatic space, generated by the stria vascularis. Indeed, the electrical energy and the ionic environment provided to OHCs are exceptional. However, a question remains as to how the OHCs can be effective at high frequencies because viscous drag increases with the frequency while the capacitive conductance of the basolateral membrane increases with frequency and significantly attenuates the receptor potential, which drives the motile mechanism in the cell body of OHCs \cite{ha1992}. 

This puzzle has been called the ``RC time constant'' problem, leading to a dispute regarding the basis for the amplifying role of OHCs: active process in the hair bundle alone \cite{Hudspeth2008}, or somatic motility coupled with hair bundle transduction \cite{lib-zuo2002,Dallos2006}, or a combination of both \cite{OMaoileidigh2010a}. The second point of view was examined by considering various mechanisms that could improve the effectiveness of somatic motility \cite{de1995,odi2003a,Iwasa2008a,Mistrik2009,Johnson2011,Meaud2011a,ospeck2012}. 

Despite their differences, all these analyses assume that the membrane capacitance, which consists of two components, linear and nonlinear, is unaffected by the mechanical load on OHCs. Of the two components, the linear component is structural, primarily based on the capacitance of the plasma membrane. Nonlinear component is due to the movement of the motile charge, which is associated with the motile function of the cell. This component has a bell-shaped membrane potential dependence in the load-free condition and the magnitude of this component at its peak can be larger than the linear capacitance \cite{a1987,sd1988}. For this reason, the motor charge appeares to enhance ``RC attenuation,'' and thereby to decrease the effectiveness of OHCs as an amplifier \cite{ha1992}.

A recent analysis, however, showed that nonlinear capacitance should depend on mechanical load, leading to a prediction that viscous drag increases mechanical energy output of OHCs by reducing the attenuation by motile charges \cite{Iwasa2016}. Here it is shown by using a simple model system that the effect of mechanical resonance is more substantial. It can fully nullify the membrane capacitance, increasing energy output of OHCs. The implications of the finding to the cochlea are discussed by connecting the input and output by additional assumptions and by examining the resulting inequality that describes the upper bound of the effectiveness of OHCs.

\section*{Model System}%%%%
Here we consider a system with an OHC, which is connected to a spring with stiffness $K$, a dashpot with friction coefficient $\eta$, and a mass $m$. 
\begin{figure}[h] %%%%%%%%%%
\begin{minipage}[c]{0.4\textwidth}
\includegraphics[width=0.9\textwidth]{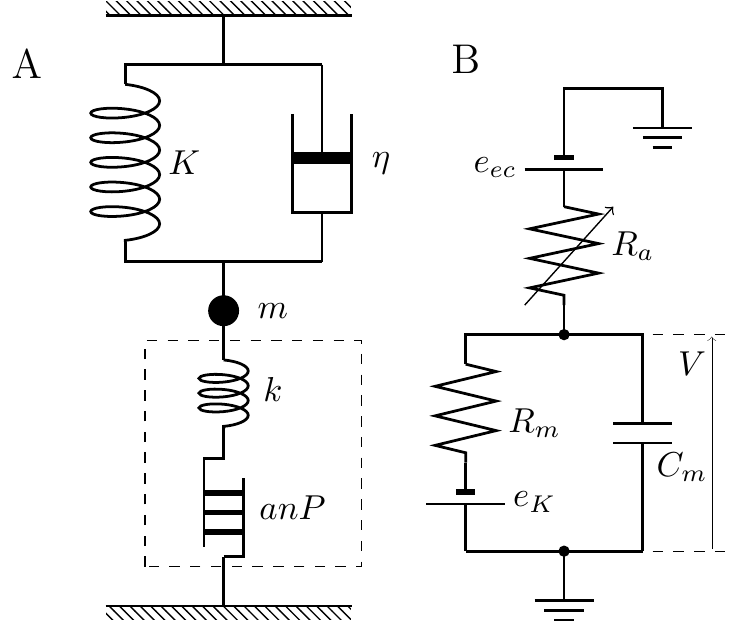}
\end{minipage}\hfill
\begin{minipage}[c]{0.6\textwidth}
\caption{\label{fig:system}\small{Mechanical connectivity and the equivalent electric circuit of the model system. The system is driven by changes in hair bundle conductance $R_a$. Unlike \emph{in vivo} condition, movement of the cell body does not affect $R_a$. In the mechanical schematics (A), $K$ is stiffness of the external mechanical load, $m$ the mass, and the drag coefficient is $\eta$. The contribution of the motile element to cell length $x$ is proportional to $anP$, where $P$, $a$, and $n$ respectively represent the fraction of the motile elements in the elongated state, unitary length change, and the number of such units, the unitary change of charge of which is $q$.  The stiffness of the cell due to the material property is $k$. The broken line indicates the border of the OHC. In the equivalent circuit (B), the membrane potential is $V$, the basolateral resistance $R_m$, and the total membrane capacitance of the basolateral membrane $C_m$, consisting of the structural capacitance $C_0$ and the contribution of charge movements in the motile element, which depends on the load. The endocochlear potential is $e_{ec}$ and the potential $e_K$ is due to K$^+$ permeability of the basolateral membrane. The apical capacitance is ignored in this model. }}
\end{minipage}
\end{figure} %%%%%%%%
We assume here that the cell has $n$ motile elements, which has two discrete states, compact and extended, and during a transition from the compact state to the extended state, the cell length increases by $a$ and the electric charge $q$ flips across the plasma membrane. The axial stiffness of the cell is $k$ (Fig.\ \ref{fig:system}). Since a set of the equations that govern this system has been derived previously~\cite{Iwasa2016}, only a brief description is given below.

\subsection*{Basic Equations}%%%
Let  $P$ be the fraction of the motile units in the extended state. Its equilibrium value $P_\infty$ follows the Boltzmann distribution,
\begin{eqnarray}\label{eq:distr}
P_\infty=\frac{\exp[-\beta\Delta G]}{1+\exp[-\beta\Delta G]},
\end{eqnarray}
with $\beta=1/(k_BT)$, where $k_B$ is Boltzmann's constant and $T$ the temperature, and 
\begin{eqnarray}\label{eq:energy}
\Delta G=q(V-V_{1/2})+\tilde{K}a^2n(P-P_0),
\end{eqnarray}
where $\tilde K=kK/(k+K)$. If the system has not reached equilibrium, $P_\infty$ can be regarded as the target value of $P$ for the given set of variables, i.e. $V$ and force applied to the motile unit, at a given moment, and changes take place toward that goal.
The equation of motion can then be expressed,
\begin{eqnarray} \label{eq:motion}
m\frac{d^2P}{dt^2}+\eta\frac{dP}{dt}&=&(k+K)(P_\infty-P),
\end{eqnarray}
which is intuitive for $m=0$. The receptor potential $V$ is described by,
\begin{eqnarray} \label{eq:potential}
\frac{e_{ec}-V}{R_a}&=&\frac{V-e_K}{R_m}+C_0\frac{dV}{dt}-nq\frac{dP}{dt}.
\end{eqnarray}
Here $R_a$ is the apical membrane resistance, which is dominated by mechanotransducer channels in the hair bundle.   The basolateral membrane has the resistance $R_m$ and the linear capacitance $C_0$, which is determined by the membrane area.

\section*{Response to Small Oscillatory Stimuli}%%%
Here we assume small periodic changes with an angular frequency $\omega$ from a resting resistance $\bar{R}_a$ of the hair bundle resistance, 
\begin{equation*}
\label{eq:ra} R_a(t)=\bar{R}_{a}(1+\hat r \exp[i\omega t]).
\end{equation*}
The response of the system should be described by small periodic changes of the variables from their steady state values:
\begin{align*}
V(t) &=\bar{V}+v \exp[i\omega t], \\
P_\infty(t) &=\bar{P_\infty}+p_\infty \exp[i\omega t], \\
P(t) &=\bar{P}+p \exp[i\omega t],
\end{align*}
where the variables expressed in lower case letters are small and those marked with bars on top are time-independent. Namely, $\bar{V}=(e_{ec}R_m+e_K\bar{R}_a)/(R_m+\bar{R}_a)$  and $\bar{P}=\bar{P_\infty}$. Thus, $\bar{P}$ is expressed by Eq.\ \ref{eq:distr} with $\Delta G$, in which $P$ is replaced by $\bar{P}$.

The equations for the small amplitudes are,
\begin{align} \label{eq:pinf-eq}
p_\infty&=-\gamma(qv+a^2n\tilde{K}p),\\ \label{eq:motion}
[-\left(\omega/\omega_r\right)^2+i\;\omega/\omega_\eta] p&=p_\infty-p, \\ \label{eq:recpot}
-\frac{e_{ec}-\overline{V}}{\overline{R}_a}\hat r &=\left(\frac 1{\overline{R}_a}+\frac 1 {R_m}\right)v+i\omega(C_0-nq\cdot p),
\end{align}
where the resonance frequency $\omega_r$, viscous roll-off frequency $\omega_\eta$, and definitions of parameters $\tilde{K}$ and $\gamma$ are given by,
\begin{align*}
\omega_r^2=(k+K)/m,\qquad \omega_\eta=(k+K)/\eta,\qquad 
\gamma=\beta\overline{P}(1-\overline{P}).
\end{align*}

Eq.\ \ref{eq:recpot} for the receptor potential can be made simpler 
by introducing two parameters,
$i_0=(e_{ec}-e_K)/(\bar{R}_a+R_m)$, and
$\sigma=1/\overline{R}_a+1/R_m$
into
\begin{align} \label{eq:simpler-rp}
-i_0 \hat r=(\sigma+i\omega C_0)v-i\omega nqp.
\end{align}

The combination of Eqs.\ \ref{eq:pinf-eq} and \ref{eq:motion} leads to,
\begin{align}\label{eq:p-by-v}
[-\left(\omega/\omega_r\right)^2+i\;\omega/\omega_\eta+(1+\gamma a^2 n\tilde{K})] p=-\gamma q v.
\end{align}
Here let us introduce a parameter $\alpha^2=1+\gamma a^2 n\tilde{K}$ for brevity. Note here that $\alpha=1$ in the absence of external elastic load and otherwise $\alpha>1$. 
\begin{figure}[h] %%%%
\begin{center}
\begin{minipage}[t]{7cm}A \end{minipage} \begin{minipage}[t]{6cm}B \end{minipage}\\
\includegraphics[width=7cm]{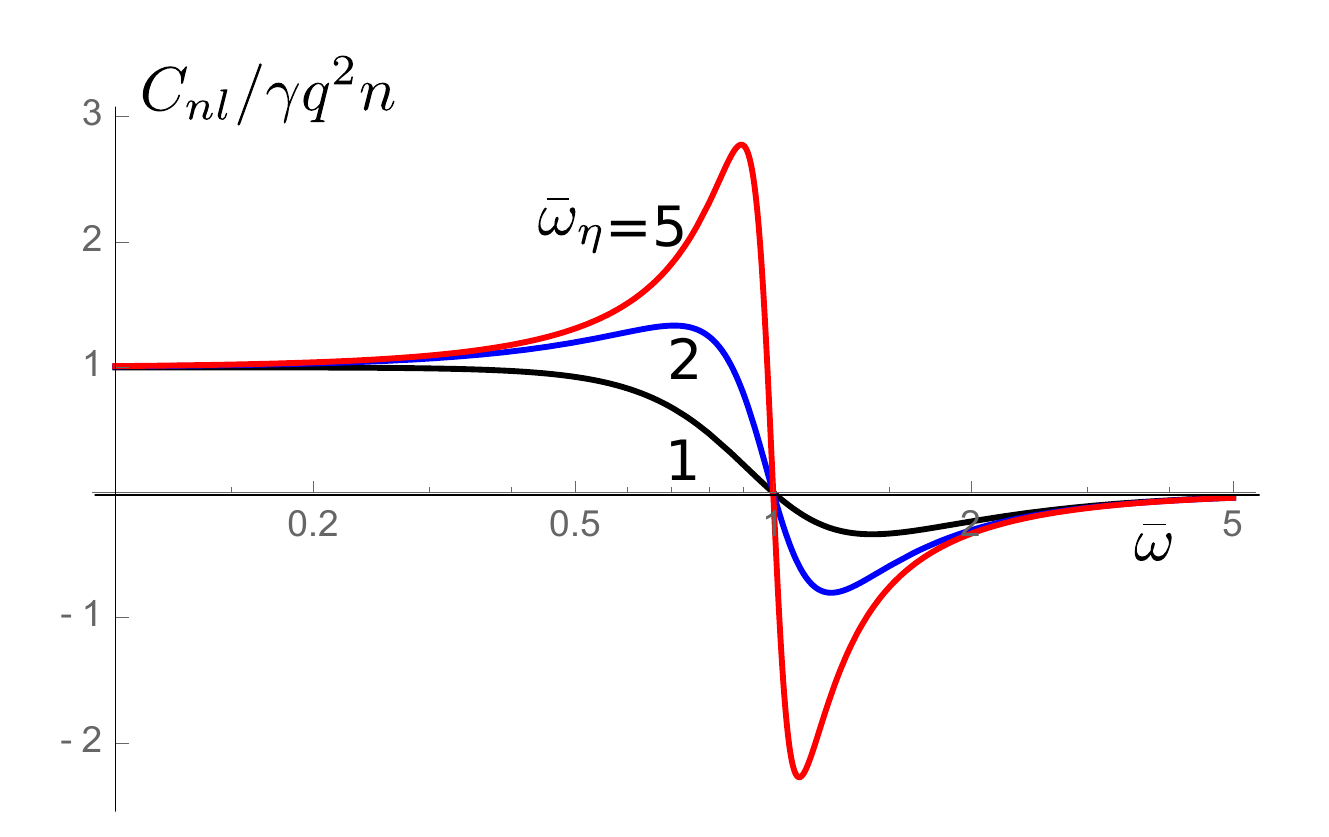}\ \ \includegraphics[width=7cm]{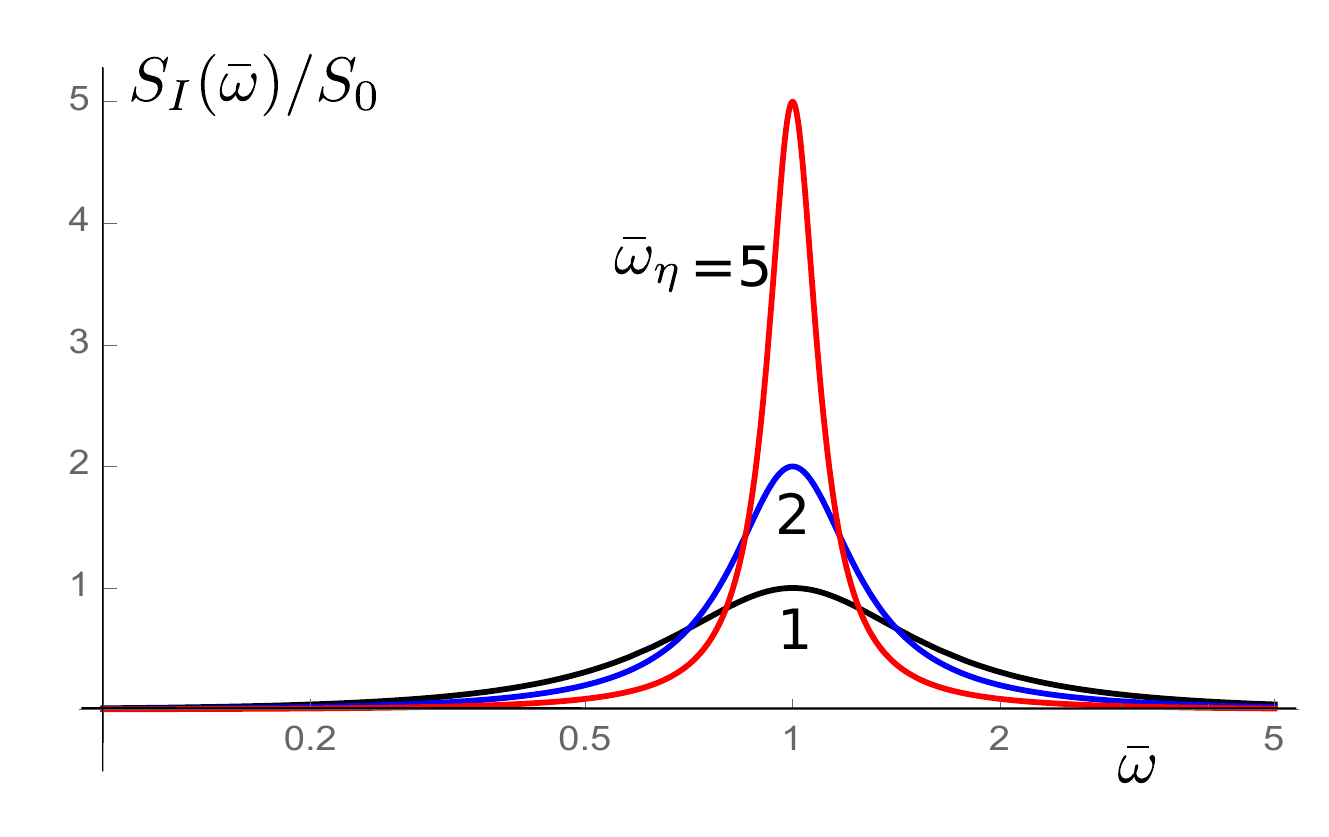}
\caption{\small{Nonlinear capacitance $C_{nl}$ and power spectral density $S_I(\omega)$ of current noise. \textbf{A:} Nonlinear capacitance plotted against $\bar\omega(=\omega/\omega_r)$. Nonlinear capacitance $C_{nl}$ is normalized by $\gamma n q^2$.  \textbf{B:} Power spectral density of current noise is plotted against $\bar\omega$. $S_I(\omega)$ is normalized by $S_0(=4\bar P(1-\bar P)q\omega_r)$. Traces respectively correspond to the values of $\bar\omega_\eta$, 1 (black), 2 (blue), and 5 (red).}}
\label{fig:cap}
\end{center}
\end{figure} %%%%%%
The contribution $C_{nl}$ of the motor charge to the membrane capacitance $C_m$ is given by $C_{nl}=(qn/v)Re[p]$. This leads to,
\begin{align}\nonumber
C_m&=C_0+C_{nl}, \\
C_{nl}&=\frac{\gamma n q^2[\alpha^2-\bar\omega^2]}{[\alpha^2-\bar\omega^2]^2+(\bar\omega/\bar\omega_{\eta})^2},
\label{eq:Cm-mres}
\end{align}
where $\bar\omega=\omega/\omega_r$,  $\bar\omega_\eta=\omega_\eta/\omega_r$, and $C_0$ is the regular membrane capacitance proportional to the membrane area of the cell.
Eq.\ \ref{eq:Cm-mres} indicates that $C_{nl}^{(max)}=\gamma q^2n$ in the absence of mechanical load, consistent with earlier studies \cite{s1991,i2010,Iwasa2016}.

Voltage oscillation $v\exp[i\omega t]$ generates current $i\omega p\exp[i\omega t]$. The admittance is given by $Y(\omega)=i\omega p/v$. Since the fluctuation-dissipation theorem \cite{Nyquist1928,Kubo1957} relates the admittance to the power spectrum of current noise with the formula  $S_I(\omega)=4k_BT\; Im[Y(\omega)]$, we have
\begin{align}
 S_I(\omega)=\frac{4\bar P(1-\bar P)q\cdot\omega_r/\bar\omega_\eta\cdot\bar\omega^2}{[\alpha^2-\bar\omega^2]^2+(\bar\omega/\bar\omega_{\eta})^2}.
\end{align}
It has a peak $4\bar P(1-\bar P)q\omega_\eta$ at $\bar\omega^2=\alpha^2$ (Fig.\ \ref{fig:cap}B). This spectral shape is quite different from that without mechanical resonance, which has high-pass characteristics \cite{i1997,Dong2000}.

Now let us examine power output elicited by hair bundle stimulation. Since the voltage change $v$ is the result of a change $r$ in the hair bundle resistance as described by Eq.\ \ref{eq:simpler-rp}, it is expressed by,
\begin{align}\label{eq:v-by-rp}
v=\frac{-i_0\hat r+i\omega nqp}{\sigma+i\omega C_0}.
\end{align}
By combining Eqs.\ \ref{eq:p-by-v} and \ref{eq:v-by-rp}, we obtain,
\begin{align}\label{eq:p-by-r}
\left[-\left(\frac{\omega}{\omega_r}\right)^2+i\omega\left(\frac{1}{\omega_\eta}+\frac{\gamma nq^2}{\sigma+i\omega C_0}\right)+\alpha^2\right] p=\frac{\gamma q i_0}{\sigma+i\omega C_0}\hat r.
\end{align}

\subsection*{High Frequency Asymptote}\label{section:hfasymptote}

Since we are interested is in high frequency range, we may assume $\sigma+i\omega C_0 \rightarrow i\omega C_0$.
Then Eq.\ \ref{eq:p-by-r} can be simplified into 
\begin{align}\label{eq:hi-asympt}
\left[-\left(\frac{\omega}{\omega_r}\right)^2+i\frac{\omega}{\omega_\eta}+\alpha^2+\zeta \right] p=-i\frac{\gamma q i_0}{\omega C_0}\hat r,
\end{align}
with a parameter $\zeta=\gamma q^2n/C_0$, which is the ratio of maximal nonlinear capacitance to the regular capacitance.

The work against drag per half cycle is 
 $E_d=(1/2) \eta \omega (\tilde K/K)^2|nap|^2$.
Power output $W_d$, which is $2\omega/(2\pi) E_d$, can be expressed by 
\begin{align} 
W_d \label{eq:Wd-HiF-Res}
= \frac{(\gamma a n q i_0)^2}{[\alpha^2 +\zeta-\bar\omega^2]^2+(\bar\omega/\bar\omega_\eta)^2}\cdot\frac{\eta k^2 \hat r^2}{2\pi (k+K)^2C_0^2},
\end{align}
using a reduced frequency $\bar\omega=\omega/\omega_r$, and $\bar\omega_\eta=\omega_\eta/\omega_r$.

\begin{figure}[h] %%%%%
\begin{center}
\begin{minipage}[t]{7cm}A \end{minipage} \begin{minipage}[t]{6cm}B \end{minipage}\\
\includegraphics[width=7cm]{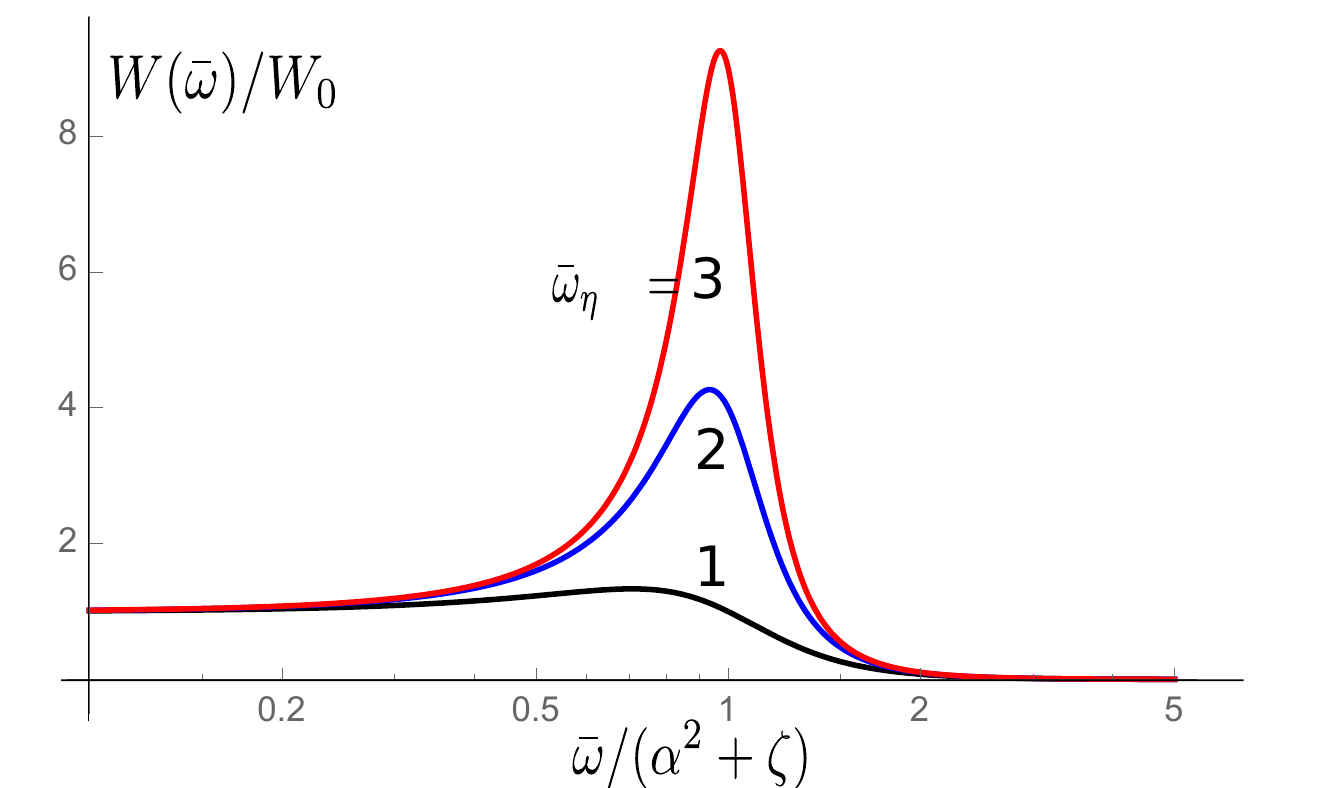}\ \ \includegraphics[width=7cm]{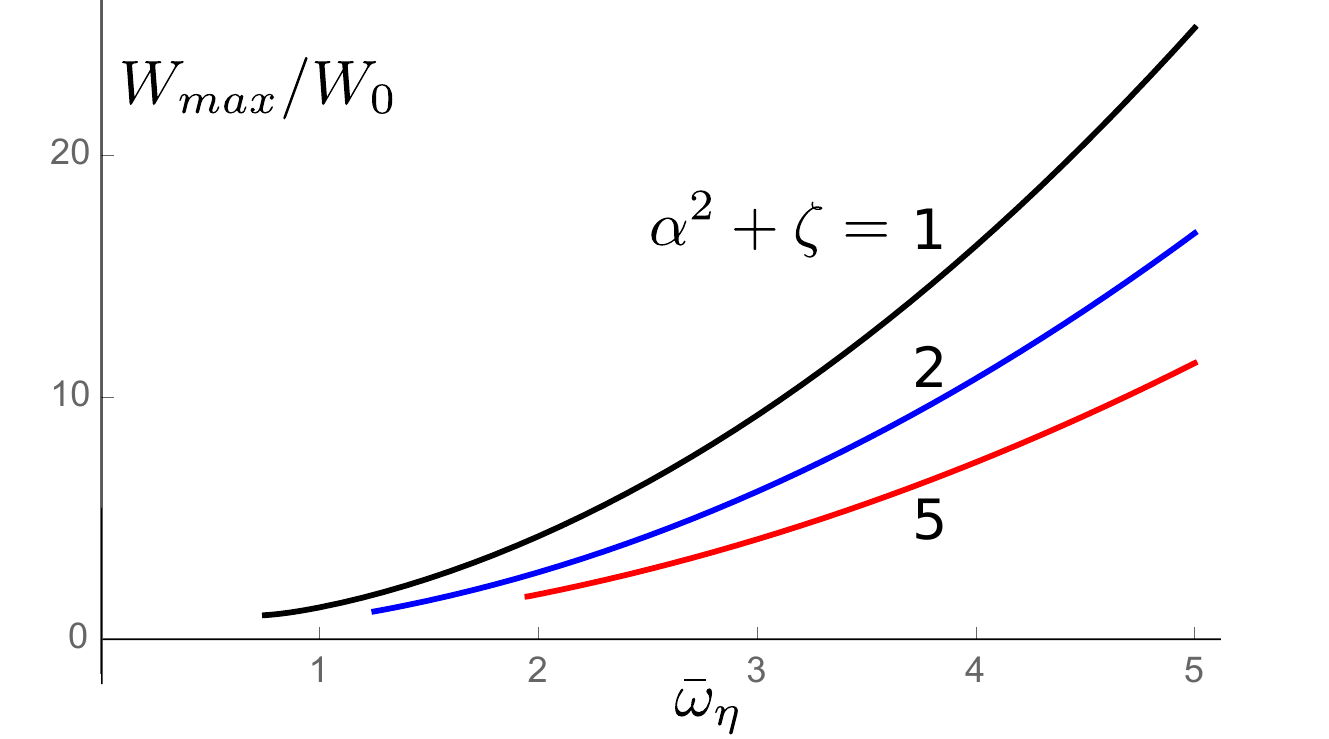}
\caption{\small{Power output per unit resistance change $(\hat r=1)$. A: Frequency dependence of power output, assuming $\alpha^2+\zeta=1$. Power output $W(\bar\omega)$ is normalized by $W_0=(\gamma a n q i_0)^2\eta k^2/[2\pi (k+K)^2C_0^2]$.  Traces correspond to the values of $\bar\omega_\eta$: 1, (black); 2, (blue); and 3 (red). B: Maximum power output plotted against $\bar\omega_\eta$. The scale of power output is the same as in A. Traces correspond to the values of $\alpha^2+\zeta$: 1, (black); 1.5, (blue); and 2 (red). }}
\label{fig:power}
\end{center}
\end{figure} %%%%%

Eq.\ \ref{eq:Wd-HiF-Res} is maximized at $\bar\omega^2=\alpha^2+\zeta-1/(2\bar\omega_\eta^2)$  and the maximal value is,
\begin{align} \label{eq:Wd-HiF-max}
W_d^{(max)}=
\frac{4(\gamma a n q i_0)^2\bar\omega_\eta^4}{4(\alpha^2+\zeta)\bar\omega_\eta^2-1}\cdot\frac{\eta k^2 \hat r^2}{2\pi (k+K)^2C_0^2}.  
\end{align}

Notice that the maximum $W_d^{(max)}$ is a monotonically increasing function of $\bar\omega_\eta$ because $\alpha^2+\zeta-1/(2\bar\omega_\eta^2)>0$. If $\bar\omega_\eta$ is sufficiently large to satisfy  $4(\alpha^2+\zeta)\bar\omega_\eta^2 \gg 1$, it can be approximated by
\begin{align} \label{eq:Wd-max-approx}
W_d^{(max)}&\approx
\frac{(\gamma a n q i_0)^2\bar\omega_\eta^2}{\alpha^2+\zeta}\cdot\frac{\eta k^2 \hat r^2}{2\pi (k+K)^2C_0^2}.
\end{align}

\subsection*{Negative Capacitance}
To understand the condition for maximizing power output, it would be instructive to examine the value of the membrane capacitance $C_m$. 

The condition for maximizing power output, $\bar\omega^2=\alpha^2+\zeta-1/(2\bar\omega_\eta^2)$ with $\bar\omega_\eta^2 \gg 1$ makes nonlinear capacitance negative:
\begin{align}
 \frac{C_0+C_{nl}}{C_0} \approx -\frac{2\alpha^2-\zeta}{2\zeta^2}\frac{1}{\bar\omega_\eta^2}.
\end{align}
This means that, under this condition, nonlinear capacitance $C_{nl}$ cancels out the regular capacitance $C_0$. The net membrane capacitance $C_m$ is small in magnitude because of the factor $1/\bar\omega_\eta^2$ and likely negative because $\alpha^2>1$ by definition and experimental data usually show $\zeta \lessapprox 2$.

\subsection*{Receptor Potential}
The amplitude of the receptor potential $v$ can be obtained from Eqs.\ \ref{eq:p-by-v} and \ref{eq:p-by-r} by eliminating $p$. At high frequencies, this combination leads to
\begin{align}\label{eq:rp}
 v=i\frac{i_0\hat r}{\omega C_0}\cdot\frac{\alpha^2-\bar\omega^2+i\bar\omega/\bar\omega_\eta}{\alpha^2+\zeta-\bar\omega^2+i\bar\omega/\bar\omega_\eta},
\end{align}
which consists of two factors. The first factor $i_0/(\omega C_0)$ can be expressed as $(e_{ec}-e_K)/[\omega(\bar R_a+R_m)C_0]$, recalling the definition of $i_0$. It indicates low-pass attenuation with a time constant $\tau_{RC}=(\bar R_a+R_m)C_0$. The second factor represents enhancement near $\bar\omega^2\approx \alpha^2+\zeta$. The magnitude of $v$ at the peak frequency can be expressed,
\begin{align}
 |v|_{max}\approx \frac{\bar\omega_\eta}{\omega_r\tau_{RC}}\cdot\frac{\zeta}{\alpha^2+\zeta}\cdot(e_{ec}-e_K)\hat r,
\end{align}
assuming $\bar\omega_\eta\gg 1$. The first factor on the right-hand-side can be expressed in a more symmetric form: $\omega_\eta\omega_{RC}/\omega_r^2$. This expression illustrates the presence of attenuation, which is still determined by $C_0$, even though the net membrane capacitance $C_m$ is virtually eliminated by the motile charge.

\section*{Frequency Limit}
Here the results obtained for our simple model system (Fig.\ \ref{fig:system}) are examined for implications to the mammalian cochlea, a complex system, specifically with regard to the limit of the effectiveness of OHCs for amplifying the oscillation in the cochlea. 

This examination is based on two major additional assumptions as in a previous treatment  \cite{odi2003a}: that the output of OHC feeds back to hair bundle displacement and that the major source of the drag is the shear in the gap between the tectorial membrane and the reticular lamina, which is essential for hair bundle stimulation. 

Hair bundle stimulation gives rise to changes $\hat r$ in hair bundle resistance, which leads to the amplitude $x$ of cell displacement, which is expressed as $x=anp\cdot k/(k+K)$ and $p$ is described by Eq.\ \ref{eq:p-by-r}. If the resulting cell displacement brings about the mechanical stimulation same as the initial one, and their phases match  \cite{odi2003a}, the movement of the system is self-sustaining. The amplitude is determined by the nonlinearity of the system \cite{odi2003a}, which is not described here.

Let us assume that hair bundle displacement $z$ and OHC displacement $x$ are proportional and described by $z=\lambda x$. The dependence of the change $\hat r$ in hair bundle resistance on hair bundle displacement $z$ has been experimentally studied. Let $g$ the sensitivity of the hair bundle transducer. Although the relationship between $z$ and $\hat r$ is nonlinear, let $g$ the mechanosensitivity at the operating point. Then a condition for an effective amplifier is given by 
\begin{align}\label{eq:f-limit}
 g\lambda |x|_{(max)}\ge\hat r,
\end{align}
where $|x|^2$ is expressed by Eq.\ \ref{eq:p-by-r} for high frequencies,
\begin{align}
 |x|^2=\frac{(\gamma a q n i_0)^2}{(\omega_r C_0)^2}\left(\frac{k}{k+K}\right)^2 H(\bar\omega) \cdot\hat r^2,
\end{align}
with
\begin{align}
 H(\bar\omega)&=\frac{1}{\bar\omega^2[(\alpha^2+\zeta-\bar\omega^2)^2+(\lambda\bar\omega/\bar\omega_\eta)^2]},
\end{align}
where $\lambda$ appears in the denominator because it changes the amplitude of length and in effect changes the drag coefficient in the subtectorial space, where the dominant drag loss is expected. Notice here that the regular capacitance $C_0$ remains as an important factor that determines the effectiveness of OHC, even though $C_m$ is very small under this condition.

\begin{figure}[h] %%%%%
\begin{center}
\begin{minipage}[t]{7cm}A \end{minipage} \begin{minipage}[t]{5.5cm}B \end{minipage}\\
\includegraphics[width=7cm]{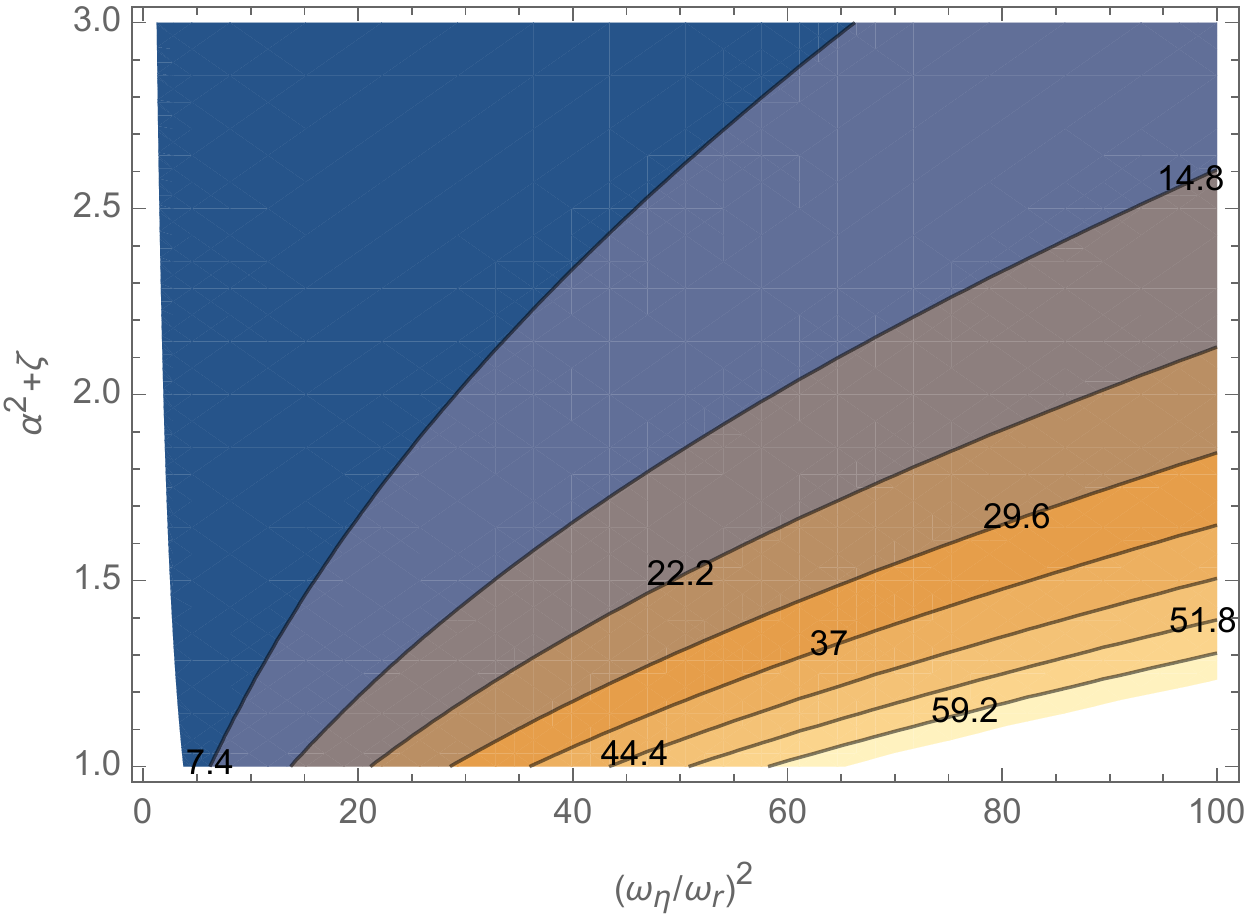}\ \ \includegraphics[width=7cm]{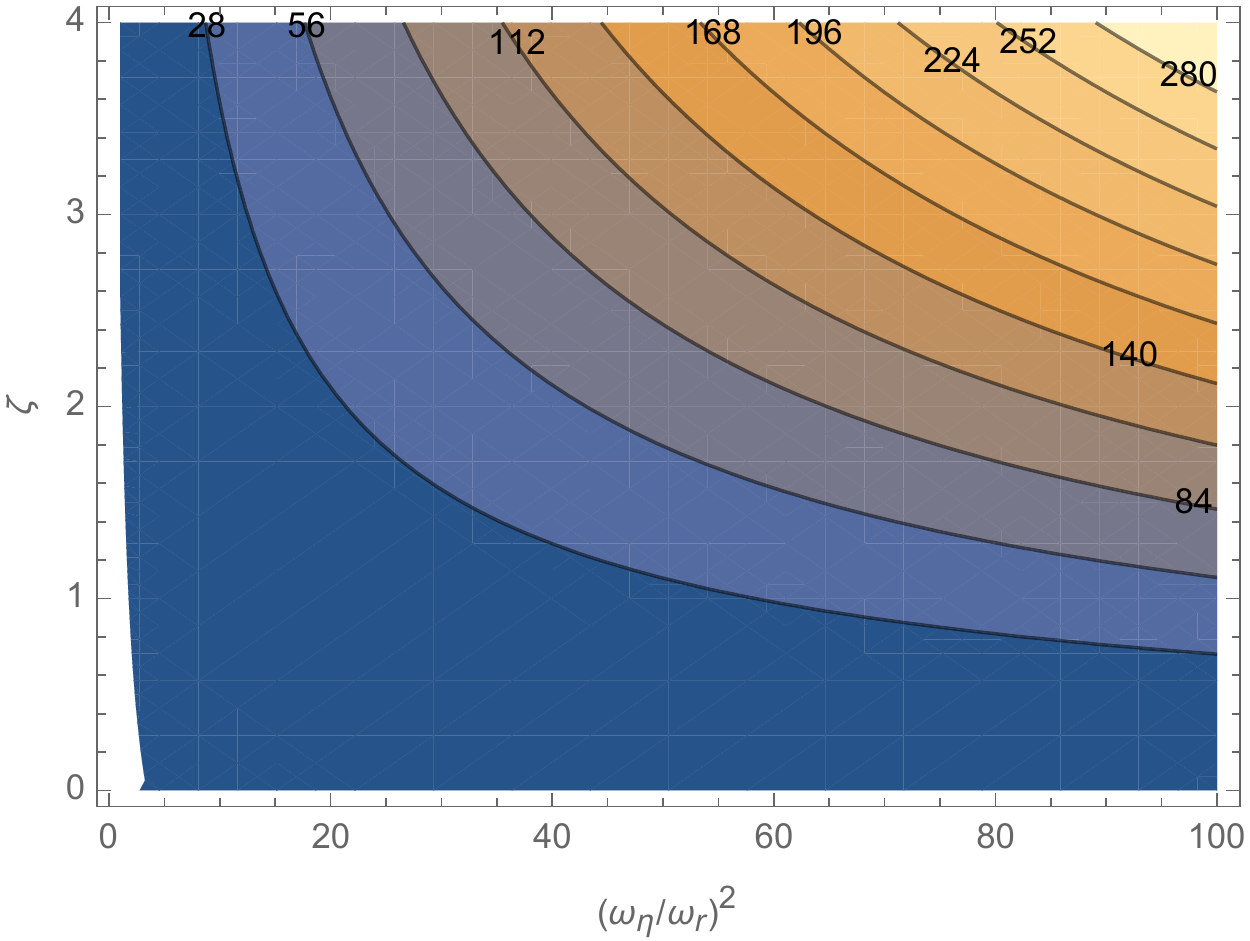}
\caption{\small{Contour plots of $H_{max}$ and $\zeta^2H_{max}$ for $\lambda=1$. A: Contour plot of $H_{max}$ Ordinate axis: $\bar\omega_\eta^2=(\bar\omega_\eta/\bar\omega_r)^2$; abscissa: $\alpha^2+\zeta$. The values of $H_{max}$ are indicated in the plot. Brighter shades indicate higher values.  B: Contour plot of $(\alpha^2+\zeta)\zeta^2H_{max}$ assuming $\alpha^2=1.1$, corresponding to a 10 kHz cell (see text). Ordinate axis: $\bar\omega_\eta^2$; abscissa: $\zeta$. The values of $(\alpha^2+\zeta)\zeta^2 H_{max}$ are indicated in the plot. Brighter shades indicate higher values. }}
\label{fig:H-max}
\end{center}
\end{figure}

Function $H(\bar\omega)$ is a monotonically decreasing function of $\bar\omega^2$ except for where the condition $(2-\sqrt 3)(\alpha^2+\zeta)<1/\bar\omega_\eta^2<(2+\sqrt 3)(\alpha^2+\zeta)$ is satisfied.  A narrow band of parameter values within this condition, the  maximum of $H$ exceeds 40 (Fig.\ \ref{fig:H-max}). This condition enables amplifying function at high frequencies. 

If the transfer function $g(z)$ is linearized to $\hat r=gz$ in the immediate neighborhood of the operating point, the frequency limit $\omega_b$ is  expressed by,
\begin{align}\label{eq:omega-r}
 \omega_b^2<(\alpha^2+\zeta) \zeta^2\left(\lambda g i_0\cdot\frac{a}{q}\cdot\frac{k}{k+K}\right)^2 H_{max}(\alpha^2+\zeta,\bar\omega_\eta/\lambda),
\end{align}
where the best frequency $\omega_b$ is related to the mechanical resonance frequency $\omega_r$ by $\omega_b^2=(\alpha^2+\zeta)\omega_r^2$ and the definition of $\zeta$ is used to replace $C_0$.  The local maximum of $H(\bar\omega)$ is expressed by $H_{max}(\alpha^2+\zeta,\bar\omega_\eta/\lambda)$. The dependence of this function on the two parameters is plotted as a contour graph (Fig.\ \ref{fig:H-max}).

 %%%%%

The inequality indicates the importance of the ratio $k/(k+K)$. The optimum condition is $K=0$ and $\omega_r^2=k/m$. While a larger value of $K$ elevates the mechanical resonance frequency $\omega_r$, it reduces $\omega_b$, making the effectiveness of higher frequency unfavorable. This issue will be discussed later.

\subsection*{Values of the Parameters for a 10 kHz cell}
Here parameter values are examined using a set of data available for 10 kHz cells. The first to be examined is the validity of an inequality $\omega_\eta/\omega_r\gg 1$, the optimizing condition for $H(\bar\omega)$. Assume that the source of the major drag is shear in the gap between the reticular lamina and the tectorial membrane. Then the drag coefficient $\eta$ is given by $\eta=\mu S^2/d$, where $\mu$ is the viscosity of the fluid, $S$ the surface area, and $d$ the gap. If $S=10\mu\mathrm{m}\times 20\mu\mathrm{m}$, and $d=1\mu$m, $\eta=1.6\times 10^{-7}$N/m, using the viscosity of water ($\mu=8\times 10^{-4}$ Pa).

Given the experimentally determined axial elastic modulus of 510 nm/unit strain \cite{ia1997}, a 20 $\mu$m long OHC has stiffness $k$ of $2.6\times 10^{-2}$ N/m (510 nm/20$\mu$m). Even if we let $K=0$, i.e. without an external elastic load, we obtain $\omega_\eta=(k+K)/\eta\approx 1.5\times 10^6$, much higher than the auditory frequency. This value would be even larger for shorter cells of higher frequency region. Thus the condition $\omega_\eta/\omega_r\gg 1$ holds.

Now let us examine the frequency limit. For a 20 $\mu$m long cell, typical of the 10 kHz, the linear capacitance is $C_0=8$ pF and $an=1\; \mu$m, which is 5 \% of the cell length. Most \emph{in vitro} experiments show the unitary motile charge of $q=0.8\;e$, where $e$ is the electronic charge. The maximal value of $\gamma$ is $1/(4k_BT)$ when the transducer channel is half open. Here $k_B$ is the Boltzmann constant and $T$ the temperature.  The resting basolateral resistance is 7 M$\Omega$ and the resting membrane potential of $-50$ mV requires the resting apical resistance of 30 M$\Omega$ \cite{Johnson2011}. These values lead to $i_0=4$ nA. 

It has been pointed out that values for the sensitivity $g$ of hair bundles determined by \emph{in vitro} experiments  tend to be underestimates due to the matching of the force probe with hair bundles \cite{Nam2015}. For this reason, $g$=1/(25 nm) \cite{rrc1986} is taken. %The largest \emph{in vitro} value is 1/(100 nm) \cite{Nam2015}.
%\\
%\hrule %%%%%%%%%%%%%%%

If we assume  $k/(k+K)=1/10$ together with $\lambda=1$ and $H=20$, an underestimate (See Fig.\ \ref{fig:H-max}A), we obtain $f_b=\omega_b/2\pi<1.1\times 10^3$, consistent with the location of 10 kHz. Power output can be evaluated using this set of parameters. With this set of the parameter values, a typical value for maximal power output would be 0.1 fW for $\hat r=0.1$. An extrapolation to the maximal output is 10 fW. These values are in a reasonable agreement with the expected output range of a single 10 kHz cell estimated from cochlear mechanics  \cite{Wang2016}.

It should be noticed, however, that these agreements do not mean that the given value for $k/(k+K)$ is reasonable as will be seen in the next section.  It simply means $k/(k+K)>0.1$ for the given set of parameters because we assumed that OHC output is at the phase optimal for amplifying to derive the inequality. %\\

% \hrule %%%%%%%%%%%%%%%%
 
\subsection*{Performance at Higher Frequencies}
For an OHC effective at higher frequencies, two conditions should be met. One is that the mechanical resonance frequency $\omega_r(=\sqrt{m/(k+K)}\;)$ must be high. The other is $\omega_b$, which is proportional to $k/(K+K)$ must be larger than $\omega_r$. For this reason if $k/(K+K)=0.1$ for a 10 kHz cell, an OHC cannot be effective at higher frequencies, as shown in the following.

The membrane resistance decreases about 3-fold for 10-times higher frequency \cite{Johnson2011}. A 3-fold reduction of membrane resistance alone would lead to a 3-fold increase in the limiting frequency. Now a 10-fold increase of $\omega_r$ requires 1 100-fold increase of the ratio $(k+K)/m$. Considering that each OHC is held by stiff Deiters' cup in at the base around the nucleus, we can expect a 10-fold difference in the stiffness $k$ between a 5 $\mu$m cell and a 20 $\mu$m cell, the elastic modulus of OHCs being approximately constant \cite{ia1997}. In addition, a 10-fold increase in the frequency reduces the thickness of boundary layer by $1/\sqrt 10$-fold. This factor may lead to factor $\sim 3$ in reducing the mass $m$. Thus, a $\sim 30$-fold increase in $k/m$ could be expected. If we assume the ratio $k/(k+K)$ is 0.1 for mechanical resonance of 10 kHz, the ratio turns into 0.03 and the limiting frequency cannot significantly exceed 10 kHz. 

If we assume, however, that the resonance at 10 kHz is achieved without the external elastic load, we require $K>2.3k$ to achieve 100-fold increase in $(k+K)/m$. This gives the maximal value of 0.3 for the stiffness ratio $k/(k+K)$, leading to a limiting frequency well above 100 KHz, even after a decrease in $H_{max}$ due to a 10-fold increase in $k$, which makes $\alpha^2\approx 2$. 

Another important factor is the ratio $\zeta$ of the magnitude of nonlinear capacitance $C_{nl}$ to the linear capacitance $C_0$. An increase in this factor can have a significant effect in elevating the limiting frequency (Fig.\ \ref{fig:H-max}B). A two-fold increase in $\zeta$ may lead to an additional 70 \% increase in the limiting frequency.  Guinea pig data indeed shows a 4-fold increase of $\zeta$ from low frequency cells ($C_0=$35 pF) to high frequency cells (5 pF) \cite{skkkt1998}. However, rat data show no significant difference between 4 kHz cells (12.1 pF) and 30 kHz cells (5.4 pF) \cite{Mahendrasingam2010}.

The inequality \ref{eq:omega-r} shows that limiting frequency is unlikely increased further by a higher value of $H_{max}$ because this factor is large where $\zeta$ is small but limiting frequency depends on $(\alpha^2+\zeta)\zeta^2H_{max}$, which is larger for large values of $\zeta$ and $\bar\omega_\eta$ (Fig.\ \ref{fig:H-max}). 

It is possible that the limiting frequency could be raised by other factors, including the amplitude ratio $\lambda$, hair bundle sensitivity $g$, or the quantity $a/q$, which may somewhat depend on the cellular property even though it can be regarded as the molecular characteristic of the motile element. 

If these factors do not significantly increase their contributions at higher frequencies, the ratio $k/(k+K)$ must remain relatively large. Since OHCs should be involved in a relative motion between the basilar membrane and the reticular lamina  \cite{Gao2014,Ren2016}, the effectiveness of OHC requires that the resonance frequency of this relative motion must be close to that of the local basilar membrane. Since the cell bodies of OHCs would be much less stiff than the basilar membrane, the associated mass must be much smaller. In this regard, experimental observations, which reveal the modes of motion in the cochlea, are of great interest to understand the detailed mechanism of the cochlear amplifier \cite{dong-oslon2013}.

\small{
%\bibliography{/Users/kuni/Dropbox/wip/bib/ohc}
%\bibliography{ohc}

}
%\pagebreak

\end{document}